\documentclass[structabstract]{aa}

\usepackage{graphicx}
\usepackage{txfonts}
\usepackage{natbib}
\bibpunct{(}{)}{;}{a}{}{,}

\begin{document}
   \title{Supernovae without host galaxies?}

   \subtitle{Hypervelocity stars in foreign galaxies}

   \author{Peter-Christian Zinn
          \inst{1}
          \and
          Philipp Grunden
          \inst{1}
          \and
          Dominik J. Bomans
          \inst{1}
          }

   \institute{Astronomical Institute, Ruhr-University Bochum, Universit\"atsstra\ss{}e 150, D-44801 Bochum\\
              \email{zinn@astro.rub.de}}

   \date{Received 05/07/2011; accepted 14/09/2011}
 
  \abstract 
   {Harvesting the SAI supernova catalog, the most complete list of supernovae (SNe) currently available, we search for SNe that apparently do not occur within a distinct host galaxy but lie a great distance (several arcmin) apart from the host galaxy given in the catalog or even show no sign of an identifiable galaxy in their direct vicinity.}
   {We attempt to distinguish between two possible explanations of this host-lessness of a fraction of reported SNe, namely (i) that a host galaxy is too faint (of too low surface brightness) to be detected within the limits of currently available surveys (presumably a low surface brightness galaxy) or (ii) a hypervelocity star (HVS) is the  progenitor of the SN that exploded kiloparsecs away from its host galaxy.}
   {We use deep imaging to test the first explanation. If no galaxy is identified within our detection limit of $\sim$\,27\,mag\,arcsec$^2$, which is the central surface brightness of the faintest known LSB galaxy so far,  we discard this explanation and propose that the SN, after several other checks, had a hypervelocity star progenitor. We focus on observations for which this is the case and give lower limits to the actual space velocities of the progenitors, making them the first hypervelocity stars known in galaxies other than our own Milky Way.}
   {Analyzing a selected subsample of five host-less SNe, we find one, SN\,2006bx in UGC\,5434, is a possible hypervelocity progenitor category with a high probability, exhibiting a projected velocity of $\sim$800\,km\,s$^{-1}$. SN\,1969L in NGC\,1058 is most likely an example of a very extended star-forming disk visible only in the far-UV, but not in the optical wavebands. Therefore, this SN is clearly due to \it{in situ} \rm{star formation. This mechanism may also apply to two other SNe that we investigated (SN\,1970L and SN\,1997C), but this cannot be determined with certainty. Another SN, SN 2005\,nc which is associated with a gamma-ray burst (GRB\,050525), is a special case that is not covered by our initial assumptions. Even with deep} \it{Hubble} \rm{Space Telescope data, a host galaxy cannot be unambiguously identified.}}
{}


   \maketitle

\section{Introduction}
\label{intro}
Contemporary supernova surveys such as the Robotic Optical Transient Search Experiment \citep[ROTSE,][]{ROTSE}, the Lick Observatory Supernova Search \citep[LOSS,][]{LOSS} or the Catalania Real-Time Transient  Survey \citep[CRTS,][]{CRTS} enable astronomers to study the nature of supernova explosions in a statistically valid way, producing more than one SN discovery per day in a variety of stellar environments. The largest catalog of SN explosions detected within these surveys is the Sternbarg Astronomical Institute (SAI) supernova catalog \citep{SAI}, which currently contains about 6000 individual SNe. It lists both properties of the SN itself, such as apparent magnitudes in various passbands and of course the type of explosion, if known, as well as details about the galaxy the supernova took place in. Therefore, the SAI SN catalog is a unique tool for studying the connections between supernovae and their host galaxies. This connection is of particular interest just because supernova properties itself may depend on their host environment as it has been extensively investigated especially for type Ia supernovae \citep[see e.g.][]{Sullivan2010,Lampleitl2010,Neill2009} because of their importance as cosmological standard candles. However, core-collapse SNe have been the topic of statistically significant studies investigating their properties with respect to their hosts \citep[e.g.][]{Hakobyan2008}. On the other hand, galaxy evolution studies can also be complemented by investigations concerning supernova explosions. For example, \cite{Neill2010} showed that the hosts of very luminous supernovae preferentially have low densities and very blue colors the latter corresponding to high specific star formation rates sSFR, which is the star formation rate divided by the stellar mass of a galaxy. From that,  they drew consequences for these galaxies in a way that their low density and low metallicity enables them to harbor such extreme SNe, implying that wind-driven mass loss prevents very luminous SNe to arise in higher mass, higher metallicity hosts. 

With this series of papers, we also attempt to follow this direction of investigating extreme galaxies through supernovae that took place in them. In an accompanying paper \citep{Zinn2011}, we analyzed the host galaxy of SN\,2009Z, which was classified as a low surface brightness (LSB) galaxy. Those galaxies are particularly hard to observe because of their low surface brightness $\mu_{\mathrm{B}}\,>\,23\,$mag\,arcsec$^{-2}$ as defined by e.g. \cite{ImpeyBothun}. \cite{Zinn2011} showed that because of the type of explosion \citep[IIb, which require progenitor masses of at least $20\,M_{\odot}$, see][]{Heger2003}, high-mass star formation must also occur in these extremely faint galaxies, which contradicts our current understanding of LSBs. Constraining the star formation history of SN\,2009Z's host galaxy, they concluded that the galaxy's star formation occurred in small, distinct bursts one of which is currently ongoing (of which the progenitor of SN\,2009Z is a product), interrupted by longer quiescent phases that this preserves the LSB nature of this galaxy, as proposed e.g. by \cite{Haberzettl2007,Vandenhoek2000}.

This particular paper aims to elucidate a special subgroup of SNe: supernovae without a distinct host galaxy. Going through the SAI SN catalog, there are on the order of 100 such SNe. They have either a large separation between the actual SN site and the center of the assigned host galaxy (see Sect. \ref{obs} for a quantitative definition of ``large'') or a completely undefined host galaxy. We discuss examples of both cases in this paper. 

The first attempt to investigate SNe Ia without hosts found with ROTSE and identifying the reason for this strange behavior was made by \cite{Hayward2005}. They give two reasons why a supernova would not be directly associated with a galaxy, which both promise to deliver interesting insights into galaxy and stellar astrophysics:
\begin{enumerate}
\item The host galaxy is simply too faint to be detected within the sensitivity limits of currently available data.
\item The progenitor star was a so-called hypervelocity star. A star with a very high space velocity ($v\,\ga\,100$\,km\,s$^{-1}$) that has escaped the gravitational potential of its parent galaxy and exploded somewhere out in ``no man's land''.
\end{enumerate}
We here focus on the second possibility because of the selection criteria of the host-less SN sample we observed (see Sect. \ref{obs}). In a subsequent paper, we will investigate the first possibility in far greater detail.

Given the trivial perception that nearly all supernovae are happening in other galaxies than our own Galaxy makes the context of hypervelocity stars (HVSs) very appealing because there have been no previous studies of these peculiar stars in foreign galaxies. Although simulations conducted by \cite{Sherwin2008} to investigate the possible trajectories of HVSs coming from M\,31  found that there are $\sim10^3$ HVSs near the Milky Way (MW), the authors point out that only future astrometric or radial velocity surveys in the MW halo will possibly identify them. Another argument was made for HE 0437-5439 being ejected from the Large Magellanic Cloud (LMC) by \cite{Bonanos2008} based on its low metallicity but was ruled out by \cite{Brown2010} using new {\it Hubble} Space Telescope (HST) observations that revealed the proper motion of HE 0437-5439 with a velocity vector directly pointing away from the MW center. Therefore, we are left today with about 30 known HVSs within or originating from the Galaxy \citep{Abadi2009}. They have mostly been found using targeted radial velocity surveys \citep[e.g.][]{Brown2009b}, so their identification requires a considerable amount of telescope time.

The HVS population has to be clearly distinguished from another population of stars with also high but not so extreme velocities: runaway stars. Those two populations differ in terms of the acceleration mechanism that provided them with such high velocities: HVSs are understood to gain their extremely high velocities during a ``sling shot'' maneuver-like passage of a binary system \citep[e.g.][]{Hills1988,Perets2009,Lu2010} in the vicinity of the massive black hole (MBH) residing at the center of our Galaxy \citep{Melia2001}. In contrast, runaway stars are a more abundant phenomenon. Their acceleration can be produced by either dynamical or binary ejection \citep{Silva2011,Gvaramadze2009} or even unspecified acceleration mechanisms \citep{Bromley2009}, but have in prevalent that they normally exhibit much smaller velocities than HVSs. We note that there is still only a small chance that these might be confused observationally because studies of runaway stars in the outer parts of the Galaxy revealed the example of HD\,271791 \citep{Heber2008}, a runaway star most likely to have formed in the outskirts of our Galaxy. It could therefore be misclassified as HVS when assuming that its origin is in the Galactic center. However, since this is only a single object, we do not consider the following analysis to suffer much from a contamination by stars similar to HD\,271791.
\cite{Abadi2009} proposed another scenario for the acceleration of a star to HVS velocities, which are typically some 100\,km\,s$^{-1}$. They claim that dwarf galaxies consumed during their last pericentric passage by our MW could contribute stars with similar high velocities to the HVS population. Those stars would then be accelerated by tidal forces caused by the interaction of the two merging systems. Their hypothesis is supported by the actual distribution of HVSs in the MW halo being highly anisotropic with an overdensity located in the direction of the constellation Leo that was identified by \cite{Brown2009a}. With the observations presented here that strongly suggest that HVSs exist in other galaxies, we also try to distinguish between those two acceleration scenarios. Moreover, \cite{Teyssier2009} demonstrated through N-body simulations that there might be a fraction of ``wandering stars'' ejected from their host galaxy through the passage of a dwarf companion. They give a lower limit on the fraction of these ``wandering stars'' of 0.05\%, suggesting observations via classical novae or supernovae.

\cite{Eldridge2011} modeled scenarios in which runaway stars had been ejected during the explosion of their binary companion as either a supernova or gamma-ray burst. They compiled predictions for runaway OB stars, red supergiants, and Wolf-Rayet stars, finding that a small but non-negligible fraction of these stars could travel more than 100\,pc, depending on the type and acceleration scenario. However, since they assume the basic acceleration to be caused by the explosion of the binary companion, their models are only very roughly comparable to other observations since our HVS candidates presented below have presumably traveled a much greater distance (several kpc). 

Throughout this paper, we adopt a flat $\Lambda$CDM cosmology with $H_{\mathrm{0}} = 71$\,km\,s$^{-1}$\,Mpc$^{-1}$ and $\Omega_{\Lambda}$ = 0.73 \citep{Komatsu2010}.

\section{Sample selection, observations, and data reduction}
\label{obs}
\subsection{Sample selection}
Starting from Grunden et al. (in prep.) who systematically searched the SAI SN catalog for host-less SNe, we draw a subsample from their compilation optimized for the identification of extragalactic HVSs. Since HVSs in the MW have typical velocities of several 100\,km\,s$^{-1}$, we chose host-less SNe that are not too far away from the assigned host galaxy. This means that the projected distance between the SN site and the center of the host galaxy given in the SAI SN catalog is fairly large (some arcminutes, depending on redshift) but there is a chance that the progenitors of core-collapse SNe could travel this distance within their lifetime. A reasonable cut for such a separation between SN site and host galaxy center is $\ga$\,10\,kpc just because this is the typical extent of the stellar discs of ``normal'' spiral galaxies such as the MW \citep[see e.g. the catalog of][]{Schmidt1992}. We point out that this cut was not strictly obeyed because the parent sample of host-less SNe itself is small (``host-lessness'' is not a common phenomenon) and hence the subsample to be observed would have suffered from this. A summary of the galaxy sample investigated here is presented in Table~\ref{sample}.

\begin{table*}
   \caption[]{Summary of the galaxy/supernova sample investigated in this work. All quantities given below were taken from the NASA Extragalactic Database (NED).}
      \label{sample}
      \centering
         \begin{tabular}{lllccccccc}
            \hline
            \noalign{\smallskip}
            Galaxy & type & assoc. SN & RA (J2000)$^{\mathrm{a}}$ & DEC (J2000)$^{\mathrm{a}}$ & galaxy $V$-mag & SN mag$^{\mathrm{b}}$ & $D_{\mathrm{25}}$$^{\mathrm{c}}$ & Distance$^{\mathrm{d}}$ & linear scale \\
             & & & h:m:s & d:m:s & AB-mag & mag & arcsec & Mpc & kpc/'' \\
            \noalign{\smallskip}
            \hline
            \noalign{\smallskip}
            NGC\,1058 & SA & SN\,1969L & 02:43:30 & +37:20:29 & 11.20 & 12.8 & 90.6$\times$84.6 & 7.3 & 0.035 \\
            NGC\,2968 & I0 & SN\,1970L & 09:43:12 & +31:55:43 & 11.73 & 13.0 & 60.4$\times$46.8 & 22.0 & 0.110 \\
            NGC\,3160 & S & SN 1997C & 10:13:55 & +38:50:34 & 14.17 & 17.5 & 63.9$\times$12.9 & 97.0 & 0.460 \\
            ??? & ??? & SN\,2005nc & 18:32:32 & +26:20:23 & $>27.4$ & ??? & ??? & 2233.3 & 6.725 \\
            UGC\,5434 & SAB & SN\,2006bx & 10:05:13 & +21:27:21 & 14.21 & 18.1 & 38.5$\times$24.0 & 78.3 & 0.373 \\
            \noalign{\smallskip}
            \hline
         \end{tabular}
\begin{list}{}{}
\item[$^{\mathrm{a}}$] For all galaxies, the center coordinates are given, except for SN 2005nc where there is no host galaxy, hence the SN coordinates are provided.
\item[$^{\mathrm{b}}$] Magnitude of the supernova as given in the initial CBAT discovery notice (therefore not necessarily AB system, mostly clear filter). Note that all magnitudes are consistent with the supernova being in their respective host galaxy using the standard candle relationships from \cite{Drout2010} and \cite{Kasen2009} for type I resp. type II supernovae.
\item[$^{\mathrm{c}}$] The isophotes were taken from the SDSS database and are measured in the $r$-band (except for NGC 1058 which is not within the SDSS coverage, hence an RC3 $B$-band value is used).
\item[$^{\mathrm{d}}$] For NGC 1058, and NGC 2968, the distances were estimated using the Tully-Fisher relation. All other distances were calculated using the redshift of the corresponding galaxy and a flat $\Lambda$CDM cosmology as described in Sect.~1.
\end{list}
\end{table*}

To prevent a contamination by low-mass progenitor stars that could possibly be old stars in the halo of a galaxy, we excluded type Ia SNe from the sample. We also checked all the remaining five SN sites displayed signs of enhanced {\it in situ} star formation (see Sect. \ref{results}).

The last constraint on our sample is that the SNe happened nearby, meaning that their redshift and the redshift of their assigned hosts is $z\,\leq\,0.05$. This cut was applied to minimize the cosmological effect, foremost Tolman dimming \citep{Tolman1930,Hubble1935}, which predicts the dimming of surface brightness with increasing distance in an expanding universe. Considering only these very low redshifts also ensures that evolutionary effects do not play a significant role in all of our discussions.

\subsection{Observations and archival data}
To assign the attribute HVS to a progenitor of a host-less supernova, we try to rule out the other explanation mentioned by \cite{Hayward2005}, namely that the progenitor belonged to a yet unknown, very faint and potentially LSB galaxy. This together with the statement that at a particular SN site no evidence of {\it in situ} star formation could be found leaves a hypervelocity progenitor as the only explanation of a supernova that is host-less. 

This goal was achieved by conducting very deep images of every individual SN site trying to identify a faint LSB galaxy located at these places. Since the faintest known LSB galaxy to date, which was found in the Hubble Deep Field South \citep[HDF-S,][]{HDFS}, has a central surface brightness of $\mu_{\mathrm{B}} = 26.9\,$mag\,arcsec$^{-2}$ \citep{Haberzettl2007}, we designed our observation to ensure that the predicted surface brightness limit is $27\,$mag\,arcsec$^{-2}$. If these observations do not reveal a faint structure at the actual SN site, then the possibility that the SN belonged to a yet unknown galaxy is excluded, leaving only {\it in situ} star formation and an hypervelocity progenitor as possible explanations.

Observations of all five supernova sites were obtained with the CAFOS focal reducer \citep{CAFOS} mounted on the 2.2\,m telescope at Calar Alto Observatory (CAHA), Almeria, Spain (program H10-2.2-029). The observations were carried out in service mode during January and May 2010. Each SN site was observed for 2\,h in total, divided into ten exposures with 720\,s each. To find even very faint objects, we chose to observe with an exceptionally wide filter available at CAHA: the $BV_{\mathrm{Roeser}}$ filter \citep{Kuemmel2001}. In comparison to standard Johnson filters (e.g. Johnson $V$ with a FWHM of $\approx$\,90\,nm), the $BV_{\mathrm{Roeser}}$ filter provides a FWHM of 156\,nm and hence a 1.7 times higher throughput. In addition, a binning of two by two pixels was applied, increasing the signal-to-noise ratio by another factor of two, but reducing the spatial resolution to $0.53\arcsec$/pixel. The airmass for all observations was around 1.5, and the seeing varied between $1\,\arcsec$ and $2.5\,\arcsec$. According to the exposure time calculator of CAFOS, this setup guarantees that we can reach the required sensitivity of $27\,$mag\,arcsec$^{-2}$. This detection limit was verified during the reduction steps as explained in the next section.

To check for {\it in situ} star formation at every individual SN site, we used archival data from (i) the all-sky survey of the UV satellite observatory GALEX \citep{GALEX} and, if available, (ii) the Optical Monitor aboard the XMM-{\it Newton} \citep{XMM} X-ray observatory. These data can be exploited to identify signs of ongoing star formation: directly via the far-UV ($FUV$) images produced by GALEX. Following \cite{Madau1998}  and \cite{Kennicutt1998}, one can directly convert the 1500\,$\mathrm{\AA}$ luminosity to a star formation rate (SFR) and hence use the $FUV$ flux as a measured by GALEX as direct SFR tracer.

\subsection{Data reduction}
Data reduction was done using standard IRAF\footnote{IRAF is distributed by the National Optical Astronomy Observatory, which is operated by the Association of Universities for Research in Astronomy (AURA) under cooperative agreement with the National Science Foundation.} procedures for de-biasing, flat-fielding, and correcting for sky illumination effects. Astrometry was done using the \texttt{wcstools} suite implemented in IRAF. Co-addition of the ten individual exposures of every pointing was done using various combination algorithms to determine the one that was the most suitable for producing images with highest sensitivity to faint, extended structures. To do this, the images resulting from every combination algorithm were checked by SExtractor \citep{Bertin1996} and optimized to detect low-luminous, extended objects. We found that averaging over all exposures and neglecting the lowest and the two highest values provided the most reliable detection thresholds for the entire image. Therefore, all pointings were combined with a \texttt{minmax} rejection routine implemented in IRAF, neglecting the one lowest and two highest pixel values. However, two out of five pointings (SN 1970L and SN 2006bx) show unexpected background variations after applying the reduction steps that lead to a significant decrease in detection sensitivity ($\sim$\,0.5\,mag\,arcsec$^{-2}$). These variations, which appear as luminous arcs between bright objects across the entire image with a typically 3\% enhanced background relative to image regions unaffected by these structures, are probably caused by internal reflections in the telescope and its optics. During the analysis of these two images, since the presence in one image and the absence in the other of a possible SN host galaxy was not in doubt (serendipitously, these background structures did not affect the actual locations of the two supernovae), we chose to cope with the lowered sensitivity. The resulting images are presented in the next section.\\

To determine whether there is a formerly undetected galaxy at the location of every individual supernova, we inspected the location of the corresponding SN as given in the SAI catalog both manually by eye as well as automatically. For the automated inspection, SExtractor was used again for two runs: One across the entire image, optimized for point-source detection (meaning a 10$\sigma$ detection is required over more than four connected pixels) to identify stars in the direct vicinity of the SN site for a photometric calibration. Since our images were obtained in the $BV_{\mathrm{Roeser}}$ filter, a very broad filter approximately covering both the classical Johnson B and V passbands, we adopted the conversion formula 
\begin{equation}
B_{\mathrm{j}} = 0.0783+0.7247(B-V)-0.0672(B-V)^2+V
\label{Bj}
\end{equation}
with an accuracy of about 10\% found by \cite{Gullixson1995}, exploiting that the $BV_{\mathrm{Roeser}}$ filter is mostly similar to the $B_j$ band \citep[e.g.][]{Kuemmel2001}. We adjusted the photometric zero-point SExtractor uses for magnitude computation in an iterative process such that the extracted magnitudes agree to within the $\la$\,0.2\,mag scatter with those of the reference stars selected from the USNO-A2 catalog\footnote{The version used in this paper can be retrieved online at \texttt{http://tdc-www.harvard.edu/catalogs/ua2.html}.}. With respect to the intrinsic scatter in the conversion formula given in Eqn.~\ref{Bj}, this accuracy seems to be a reasonable value. After that, a second run was performed that was optimized for the detection of extended sources: this consisted of a 3$\sigma$ detection over more than 16 connected pixels corresponding to a diameter of $2.12\arcsec$ when arranged in a square in an area of $1\arcmin\,\times\,1\arcmin$ around the cataloged SN position to determine an accurate detection threshold (with respect to extended, faint sources) in this region of interest. This indicates that the detection limit of $\mu$=27\,mag\,arcsec$^2$ as calculated prior to the observation runs could be reached fairly well. Only for the two observation runs (SN\,1970L and SN\,2006bx) exhibiting some strange background patterns as mentioned above did the detection limit drop to $\sim\,$26.5\,mag\,arcsec$^2$.

After this automated analysis step, every image was inspected by eye to verify the SExtractor result. We had to take into consideration that a very faint galaxy is barely visible to the human eye but falls just beyond the chosen SExtractor detection requirements. The resulting findings of these steps will be discussed in detail in the next section for every individual SN site, and a summary is presented in Table~\ref{data}.

\begin{table}
   \caption[]{Summary of the CAFOS data used in this work.}
      \label{data}
      \centering
         \begin{tabular}{lccl}
            \hline
            \noalign{\smallskip}
            Supernova & $\mu_{\mathrm{thresh}}$$^{\mathrm{a}}$ & $BV_{\mathrm{lim}}$ $^{\mathrm{b}}$& remarks \\
             & mag\,arcsec$^{-2}$ & mag & \\
            \noalign{\smallskip}
            \hline
            \noalign{\smallskip}
            SN\,1969L & 26.98 & 24.88 & \\
            SN\,1970L & 26.62 & 24.74 & background variations \\
            SN\,1997C & 26.95 & 24.81 & \\
            SN\,2005nc & 26.94 & 24.92 & \\
            SN\,2006bx & 26.68 & 24.69 & background variations \\
            \noalign{\smallskip}
            \hline
         \end{tabular}
\begin{list}{}{}
\item[$^{\mathrm{a}}$] SExtractor-derived 3$\sigma$ detection limit for extended sources in $BV_{\mathrm{Roeser}}$ filter.
\item[$^{\mathrm{b}}$] SExtractor-derived 3$\sigma$ detection limit for point sources in $BV_{\mathrm{Roeser}}$ filter
\end{list}
\end{table}

\section{Method and results}
\label{results}
Before examining the individual SNe, we first introduce our technique applied here to ascertain whether a supernova's progenitor is a HVS and compare it to HVS identification methods in the Galaxy.

As explained in Sect.~\ref{intro}, the basis of our HVS identification method is to rule out other explanations of a supernova appearing to be host-less (that is an LSB host galaxy that has not yet been detected) by using ultra-deep imaging data. This creates a candidate sample of host-less SNe, or (more precisely) of supernova progenitors that owing to the absence of a galaxy at the very SN site, are most likely to be ejected from the galaxy nearest to the actual SN position. Because of the cut in angular separation between nearest galaxy and SN we applied to select our host-less SNe sample (see Sect.~\ref{obs}), this implies average velocities of the progenitor star of several hundred upto more than 1000\,km\,s$^{-1}$.

The estimate of velocity is thereby calculated as follows: given the accurate positions of both the SN and the nearest galaxy, one can easily compute the physical distance between the SN and the galaxy center. Because of the cut in redshift we applied during our sample selection, this distance could be calculated using simple Euclidean geometry. Nevertheless, we chose to use cosmologically correct angular size distances corresponding to the distance of the apparent host galaxy. Depending on whether we could use redshift-independent distances (as for SN 1969L and SN 1970L) or we had to calculate a distance utilizing the cosmology calculator of \cite{Wright2006}, we could easily compute the linear scale at the corresponding distance. Since we obviously do not know where exactly in its parent galaxy the star a started its journey, we always use the center of the host galaxy as starting point. This seems to be the most reasonable assumption just because the classical acceleration scenario is a ``sling shot manoeuvre''-like passage of the star passing the galaxy's central supermassive black hole. In addition to this distance, we need a travel time to calculate an average velocity. This travel time is approximated by the lifetime of the progenitor. This is of course a much poorer assumption than the starting point being the center of the galaxy because the star would then have had to have formed in the direct vicinity of the central SMBH where, in non-active galaxies, not much material which is sufficient for massive star formation usually is present \citep[e.g.][]{Knapen2006}. Hence, this assumption causes the travel time to become an upper limit which then causes the derived velocity to be a lower limit. The lifetime of the star can be estimated fairly well given the exact type of supernova explosion. On the basis of the studies of \cite{Heger2003} and \cite{Smartt2009a}, detailed model calculations as well as observational constraints are available that enable us to consider which star (depending on mass) ends in which supernova type. Using the stellar evolution models of \cite{Schaller1992}, this different stellar masses can then be identified with a certain lifetime. We also point out here that the lifetimes we adopted are always those of the lightest star that could create a certain supernova to preserve the character of the travel time as an upper limit. For example, \cite{Smartt2009b} showed that there are well-defined lower and upper limits for a star to explode as a type IIP supernova, the most common core-collapse supernova, which is $m_{\mathrm{min}}\,=\,8.5\pm1\,M_{\odot}$ and $m_{\mathrm{max}}\,=\,16.5\pm1.5\,{M_{\odot}}$. Following this, for the progenitor of a type IIP supernova we always uses a lifetime of 30\,Myr according to the \cite{Schaller1992} calculations for a 9\,$M_{\odot}$ mass star to create an upper limit to the lifetime. The resulting estimate of velocity then reads just as the distance between the center of the nearest galaxy and the SN position divided by the travel time of the progenitor which is approximated by the maximum possible lifetime of the progenitor star.

To compare our method of attributing a star as a HVS with commonly used methods for stars in the Galaxy, we highlight three major differences. First of all, the velocities we determine here is always the \emph{velocity component projected on the plane of the sky},which is very different from the velocities determined for HVSs in the Galaxy, which are measurements of the \emph{radial velocity component}. Moreover, Galactic HVSs are always classified based on a spectrum yielding very accurate velocities whereas the velocities derived here are coarse lower limits. This is predominantly because of the assumed mass of the progenitor, which we always chose to be the lowest possible mass for the corresponding type of explosion and can have large effects just because the lifetime (on the main sequence) is proportional to $M^{-2.5}$. This means, for example, that with our method a 16\,$M_{\odot}$ star (lifetime about 10\,Myr), which would show up as type IIP supernova would be treated as a 9\,$M_{\odot}$ star with a lifetime of 30\,Myr in our calculations. The second difference is that we calculate proper velocities in contrast to classical HVS identification methods that use the Doppler shift of a star's spectral features  to determine its velocity, which is then a radial velocity. The third difference is that we calculate velocities averaged over a star's entire life in contrast to the Doppler-shift velocities derived for Galactic HVSs, which are obviously current velocities. 

Those differences ensure that the extragalactic HVSs presented below are not exactly compareable to the HVSs in our Galaxy. If we consider the entire phenomenon as simply a star moving with an exceptionally high velocity, then the two completely differently identified populations of stars can be closely related, justifying our reference to a ``hypervelocity star'' in the examples presented below. A summary of the investigated supernovae is presented in Table~\ref{supernovae}.

\begin{table*}
   \caption[]{Summary of the results for every individual supernova event investigated in this work.}
      \label{supernovae}
      \centering
         \begin{tabular}{lllcccccl}
            \hline
            \noalign{\smallskip}
            Galaxy & assoc. SN & SN type$^{\mathrm{a}}$ & separation$^{\mathrm{b}}$ & distance$^{\mathrm{c}}$ & $M_{\mathrm{min}}$$^{\mathrm{d}}$ & $\tau$$^{\mathrm{e}}$ & velocity & remarks \\
             & & & arcmin & kpc & $M_{\odot}$ & Myr & km\,s$^{-1}$ & \\
            \noalign{\smallskip}
            \hline
            \noalign{\smallskip}
            NGC\,1058 & SN\,1969L & IIP & 3.46 & 7.3 & 8 & 30 & 238 & clear {\it in-situ} star formation \\
            NGC\,2968 & SN\,1970L & Ib/c & 2.17 & 14.3 & 15 & 13 & 1076 & formation in connecting bridge likely\\
            NGC\,3160 & SN\,1997C & Ib/c & 1.60 & 44.2 & 15 & 13 & 3327 & misclassified SN, likely Ia \\
            ??? & SN\,2005nc & Ic & ??? & ??? & ??? & ??? & ??? & GRB at $z=0.61$, no host detected \\
            UGC\,5434 & SN\,2006bx & II? & 1.15 & 26.8 & 8 & 30 & 848 & most likely to be a HVS \\
            \noalign{\smallskip}
            \hline
         \end{tabular}
\begin{list}{}{}
\item[$^{\mathrm{a}}$] As given by \cite{SAI}.
\item[$^{\mathrm{b}}$] Angular separation between the actual SN position and the center of the associated host galaxy.
\item[$^{\mathrm{b}}$] Corresponding projected distance between SN and host using the linear scale given in Table~\ref{sample}.
\item[$^{\mathrm{d}}$] Minimum mass of the progenitor star of a supernova of the given type from \cite{Heger2003,Smartt2009a}.
\item[$^{\mathrm{e}}$] Corresponding lifetime of a star with $M_{\mathrm{min}}$ as derived by \cite{Schaller1992}.
\end{list}
\end{table*}

\subsection{SN\,1969L}
\begin{figure}
\centering
\includegraphics[width=0.5\textwidth]{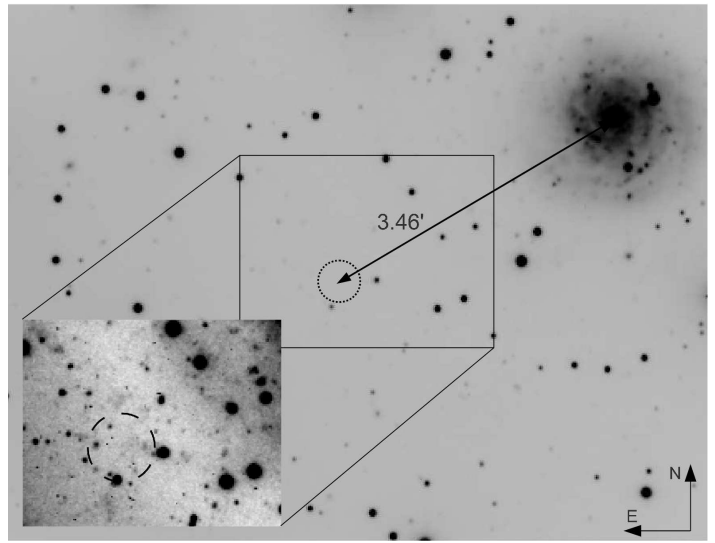}
\caption{CAFOS $BV_{\mathrm{Roeser}}$ image of the region around SN\,1969L, sized $7.7\arcmin\times6.0\arcmin$. The dashed circle indicates the exact SN position in both frames whereas the solid box indicates the borders of the small subimage in the lower left corner that shows the direct vicinity of SN\,1969L with extremely wide cut levels in order to identify even very faint structures. The arrow indicates the distance of $3.46\arcmin$ or 7.2\,kpc between the position of SN\,1969L and the center of its assigned host galaxy, NGC\,1058 (upper right).}
\label{SN1969L}
\end{figure}
Fig.~\ref{SN1969L} shows the final $BV_{\mathrm{Roeser}}$ image of our observations of the SN\,1969L (type IIP) region. As described above, ten individual exposures with 720\,s each were co-added to obtain this image with a 3$\sigma$ detection limit to a surface brightness of 26.98\,mag\,arcsec$^{-2}$, according to our validation runs using SExtractor described in Sect.~\ref{obs}. As is clearly visible in the close-up image (lower left of Fig.~\ref{SN1969L}), neither a clear galaxy-like structure nor a faint indication of a galaxy could be identified at the position of SN\,1969L by eye. For convenience, the cut levels of the close-up image are set to very wide values in order to reveal even faint, barely visible structures. This negative result is confirmed by the automated SExtractor search for faint, extended structures as SExtractor did not indicate a detection at this position. Therefore, we rule out the possibility that the progenitor of SN\,1969L was situated in a faint, yet undetected LSB galaxy and assume that the progenitor belonged to the spiral galaxy that is visible in the upper part of Fig.~\ref{SN1969L}, NGC\,1058. 

\begin{figure}
\centering
\includegraphics[width=0.5\textwidth]{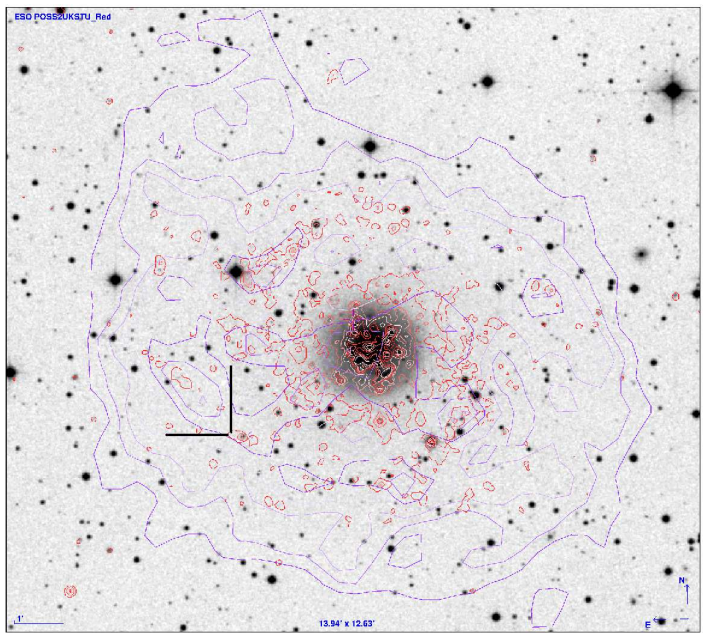}
\caption{DSS $B$-band image of NGC\,1058 with GALEX $FUV$ contours (red) and WHISP \citep{WHISP} H\,I contours (purple) overlaid. As one can easily see, the ultraviolet disk as well as the H\,I disk of NGC\,1058 are both much more ($\sim$\,3 times for the $FUV$ and $\sim$\,5 times for the H\,I disk) extended than the optically visible disk. Given the fact that SN\,1969L is located well within this $FUV$ disk (position marked with two ticks), the most likely assumption would be that its progenitor is not a HVS but a massive product of the ongoing star formation which is traced by the $FUV$ emission at this locus.}
\label{NGC1058}
\end{figure}

Since NGC\,1058 is a very nearby galaxy \citep[$z = 0.001728$,][]{Houchra1999} at a distance of only 7.3\,Mpc, the physical distance $d$ between its center and the SN position, $d = 207.6\arcsec\,\times\,0.035\,$kpc$/\arcsec = 7.3$\,kpc, is also at the lower limit of our sample (the soft selection criterion was $d\,\ga\,10\,$kpc). Hence, it is most important in this case to check for {\it in situ} star formation at this small distance from the galaxy, even though the galactic disk (which displays a high star formation activity visible as a flocculent structure of H\,II regions) seems to end quite before reaching the distance $d$ according to our deep $BV_{\mathrm{Roeser}}$ image. To check for star formation activity at the SN position, we used far-ultraviolet ($FUV$) data obtained by GALEX during its All-Sky Imaging Survey (AIS) that reaches a limiting magnitude in this band of about 20.0\,AB-mag. Fig.~\ref{NGC1058} shows the corresponding $FUV$ image overlaid as a contour plot on a DSS $V$-band image. 
As one can clearly see, the extent of NGC\,1058 in the $FUV$ is much larger (about three times in diameter) compared to the optical wavelength range. In particular, it extends to the location of SN\,1969L whose position is indicated by a cross in Fig.~\ref{NGC1058}, revealing that it is located directly within a $FUV$-luminous ``blob''. Integrating the $FUV$ emission of this blob (corresponding to 19.6\,AB-mag) and applying the calibration of \cite{Kennicutt1998} to determine a star formation rate, one infers that $SFR\,\sim\,3\cdot10^{-3}\,M_{\odot}\,\mathrm{yr^{-1}}$. Owing to this low but non-negligible star formation rate and the general shape of the star-forming ($FUV$) disk of NGC\,1058, we conclude that the progenitor of SN\,1969L has not been ejected from its parent galaxy to reach its explosion position but much more likely has been formed at this position in the very outskirts of NGC\,1058.

This interpretation is supported by other work: \cite{Kruit1984} conducted H\,I observations of NGC\,1058 using the Westerbork Synthesis Radio Telescope (WSRT), resulting in an H\,I map with a resolution of 45$\arcsec$. They traced the gaseous disk to a limiting column density of about $5\cdot 10^{19}\,$cm$^{-2}$ at a galactocentric distance of either 6.5$\arcmin$ or 13.7\,kpc. This again underlines that the optically visible disk is much smaller and is not a good tracer of the entire extent of the galaxy. Furthermore, deep H$\alpha$ observations of NGC\,1058 were conducted by \cite{Ferguson1998}, revealing the presence of H\,II regions out to two optical radii (defined by the B band isophote corresponding to 25\,mag\,arcsec$^{-2}$). On the basis of the data, they deduced a type II supernova rate in the outermost regions of the galaxy of 1-4\,SNe\,pc$^{-2}\,$Gyr$^{-1}$. For comparison, the rate of core collapse SNe in the inner parts of our Galaxy out to $\sim$\,4\,kpc given by \cite{vandenBergh1991} is on the order of 0.1\,SNe\,pc$^{-2}$\,Gyr$^{-1}$.

\subsection{SN\,1970L}
\begin{figure}
\centering
\includegraphics[width=0.5\textwidth]{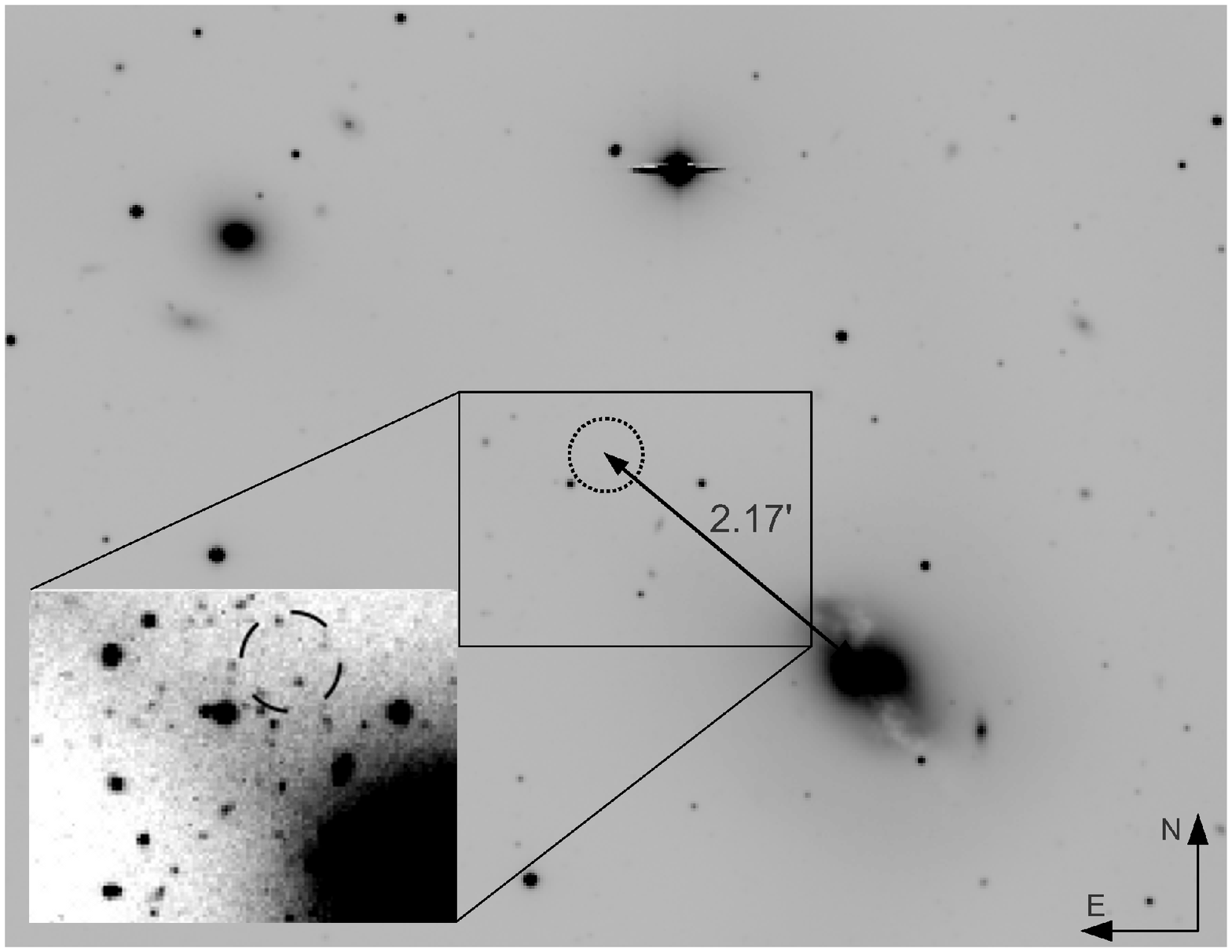}
\caption{CAFOS $BV_{\mathrm{Roeser}}$ image of the region around SN\,1970L, sized $7.7\arcmin\times6.0\arcmin$. See Fig.~\ref{SN1969L} for explanations.}
\label{SN1970L}
\end{figure}
SN\,1970L, a type Ib/c supernova discovered on October 31 by \cite{Wild1970}, shows an angular separation from the center of its assigned host galaxy, NGC\,2968, of 2.17\arcmin. Given a linear scale at a redshift $z = 0.005224$ \citep[the redshift of NGC\,2968 found by][]{diNella1995} of 0.11\,kpc/\arcsec, this converts to a physical distance of $d = 130.2\arcsec\,\times\,0.11\,$kpc$/\arcsec = 14.3$\,kpc, about two times as large as the galaxy's major axis (again defined as the isophote where the surface brightness drops to 25\,mag\,arcsec$^{-2}$). Hence, this SN is an excellent candidate to be an extragalactic HVS. Checking our deep $BV_{\mathrm{Roeser}}$ image (Fig.~\ref{SN1970L}) with a detection threshold of 26.62\,mag\,arcsec$^{-2}$ as well as the resultant source catalog from the SExtractor search, no indications of a faint LSB galaxy in the direct vicinity of SN\,1970L could be found. NGC\,2968 is also almost undetected in the corresponding GALEX $FUV$ image. Hence, SN\,1970L's progenitor might neither belong to a faint, yet undetected galaxy nor be a product of {\it in situ} star formation. 

\cite{Heger2003} found that the minimum mass of a type Ib/c progenitor star is about 15$\,M_{\odot}$, implying a lifetime of $\tau\,\approx\,$13\,Myr according to the \cite{Schaller1992} stellar evolution models. Hence, one can compute the lower limit to the progenitor's average proper velocity described in Sect.~\ref{obs} to be
\begin{equation}
v\,\geq\,d/\tau = 1076\,\rm{km\,s^{-1}}.
\end{equation}
This velocity is at the very top end of the velocity distribution of known HVSs in the Galaxy. For comparison, the fastest HVSs in the compilation of \cite{Abadi2009} shows a velocity of about 500$\,\rm{km\,s^{-1}}$. A possible explanation of this high velocity is that NGC\,2968 is engaged in a pair interaction with NGC\,2970 (the small elliptical galaxy visible in the upper left part of Fig.~\ref{SN1970L}). This becomes obvious because of a luminous bridge connecting the two galaxies \citep{Sandage1994}, which also show very similar luminosity distances (22.1\,Mpc and 22.9\,Mpc). Its total extend is about 5$\arcmin$ (from galaxy center to galaxy center) or 32.4\,kpc. This luminous bridge is also the reason for the detection threshold of our deep $BV_{\mathrm{Roeser}}$ image being about 0.5 mag less deep than theoretically calculated, because SN\,1970L is located directly within the bridge as already noted in the discovery paper by \cite{Wild1970}. Despite the bridge being clearly visible in our deep $BV_{\mathrm{Roeser}}$ image, no signs of star formation are visible in the GALEX $FUV$ images. This strongly constrains the star formation rate within the bridge to be less than 0.25$\,M_{\odot}\,\rm{yr^{-1}}$, given the GALEX AIS sensitivity limit at the distance of NGC\,2968.

All of these results together for the connecting bridge between NGC\,2968 and NGC\,2970 imply that the progenitor of SN\,1970L has not been accelerated by the common mechanism involving the presence of a SMBH but is more likely to be accelerated by the tidal interaction of the two galaxies, a mechanism very similar to that proposed by \cite{Abadi2009} and \cite{Teyssier2009}. 

\subsection{SN\,1997C}
\begin{figure}
\centering
\includegraphics[width=0.5\textwidth]{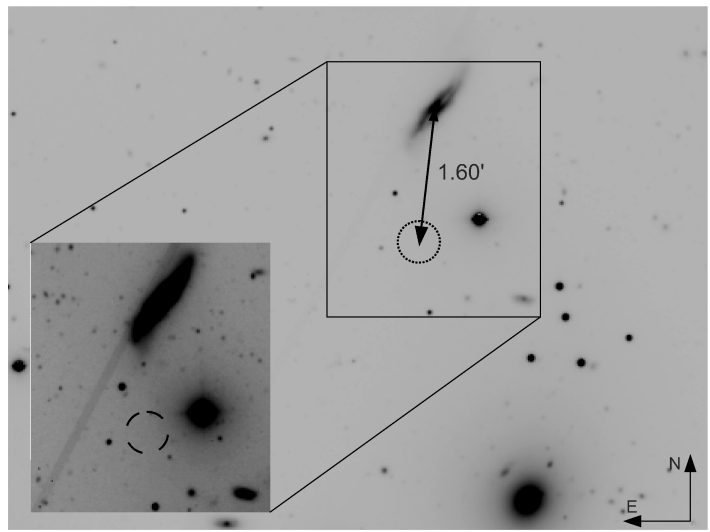}
\caption{CAFOS $BV_{\mathrm{Roeser}}$ image of the region around SN\,1997C, sized $7.7\arcmin\times6.0\arcmin$. See Fig.~\ref{SN1969L} for explanations. Note that there is a slight satellite track just at the northern side of the galaxy visible in the close-up image.}
\label{SN1997C}
\end{figure}
SN\,1997C is a type Ib/c supernova whose host galaxy is assigned to be NGC\,3160, a small spiral galaxy at redshift $z = 0.023083$ \citep{Mahdavi2004}, see Fig.~\ref{SN1997C}. Their angular separation is $1.60\arcmin$, which is equal to $d$ = 44.2\,kpc at this redshift. As for the other two SNe discussed above, no indication of a faint galaxy at the location of SN\,1997C could be found, down to a detection limit of 26.95\,mag\,arcsec$^{-2}$. There is also visible $FUV$ emission neither in the outer regions of NGC\,3160 since GALEX did not detect this galaxy within its AIS in the $FUV$ band at all, nor  at the location of SN\,1997C. Therefore, we exclude both a faint LSB galaxy to which the progenitor of the SN belonged, as well as {\it in situ} star formation as a reason for SN\,1997C happening this far away from NGC\,3160, leaving a HVS progenitor as the only explanation. We note that, as NGC\,3160 belongs to a small galaxy group, the possibility of {\it in situ} star formation could not completely be ruled out. \cite{Hatch2008} used very deep HST ACS data to demonstrate that in the galaxy cluster surrounding MRC\,1138−262 (the ``Spiderweb Galaxy'' at $z = 2.156$), nearly half of the star formation traced by $FUV$ light occurs outside any galaxy-like structure. Taking advantage of the extremely high ACS sensitivity, they could trace intergalactic light down to 27.5\,mag\,arcsec$^{-2}$ in the rest-frame $FUV$ band. Lacking similar deep $FUV$ data, we cannot exclude {\it in situ} star formation to these levels. Nevertheless, owing to cosmological surface brightness dimming, which becomes significant at these redshifts (surface brightness must be corrected by a factor of $(1+z)^4$ owing to spatial expansion in addition to decreases in both photon energy and arrival time), we can estimate that our cut for a non-detection of 27.0\,mag\,arcsec$^{-2}$ in $BV_{\mathrm{Roeser}}$ would correspond to a sensitivity of about 32\,mag\,arcsec$^{-2}$ when placed at $z = 2.156$. Therefore, we consider the {\it in situ} formation scenario for the progenitor of SN\,1997C as highly unlikely.

Again assuming a minimum progenitor mass for this type of supernova explosion of 15$\,M_{\odot}$, the progenitor lifetime is also $\tau\,\approx\,$13\,Myr, just as for SN\,1970L. Hence, the lower limit to the average proper velocity of SN\,1997C's progenitor is
\begin{equation}
v\,\geq\,d/\tau = 3327\,\rm{km\,s^{-1}}.
\end{equation}
This velocity is even higher than that derived for SN\,1970L, but may have the same cause: Revisiting the BV image of NGC\,3160, its disk appears to be significantly disturbed, exhibiting an S-shaped warp. Those warps are commonly thought to originate predominantly from interactions between two galaxies \citep[e.g.][]{Barnes1992,Reshetnikov1998}, either in the form of a massive galaxy accreting a dwarf (causing most of the low-amplitude S-shaped warps) or two massive galaxies engaged in a tidal interaction \citep[causing high-amplitude warps,][]{Ann2006}. This assumption is consistent with our findings for NGC\,3160, which is also listed as UGC\,05513 in the Uppsala General Catalog of Galaxies. In this framework, \cite{Nilson1973} highlight the disturbed shape of NGC\,3160 and assume that it is caused by an interaction with the nearby elliptical NGC 3158, which is the brightest galaxy in a small (about 30 members) galaxy group to which NGC\,3160 also belongs \citep{Sandage1994}. Therefore, an interaction between those two galaxies, resulting in the accidental ejection of SN\,1997C's progenitor, is a likely scenario.

Owing to the extremely high velocity that is necessary to transport the progenitor of SN\,1997C from its assigned host galaxy to its actual explosion site, another possibility should, however, be mentioned here. \cite{Matheson2001} indicate that SN\,1997C was misclassified because the only available spectrum of this SN is one taken several weeks after maximum light. Comparing this spectrum to other SN spectra, they found that the spectrum of SN\,1997C is most similar to that of the subluminous type Ia supernova SN\,1991bg. This would imply that the progenitor star of SN\,1997C was a low-mass star with a fairly long lifetime belonging to the halo of NGC\,3160. Since this explanation seems to be the one with the fewest assumptions, we consider this to be the most likely explanation of the large separation between SN\,1997C and its host.

\subsection{SN\,2005nc}
\begin{figure}
\centering
\includegraphics[width=0.5\textwidth]{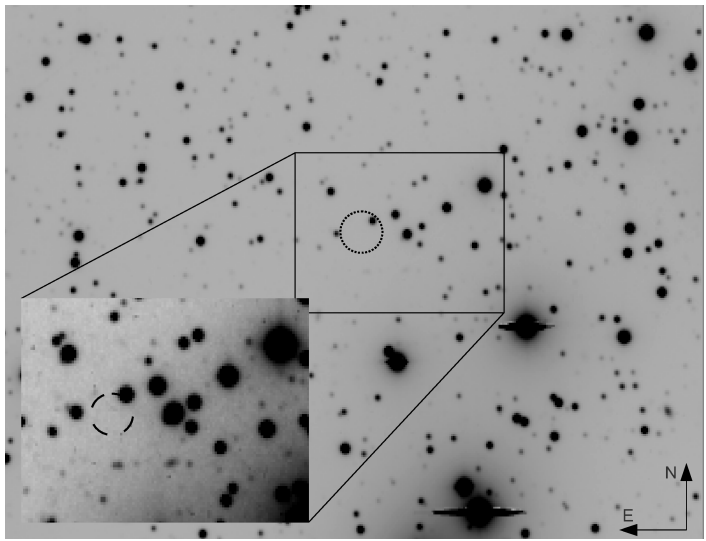}
\caption{CAFOS $BV_{\mathrm{Roeser}}$ image of the region around SN\,2005nc, sized $6.7\arcmin\times5.0\arcmin$. Because there is absolutely no galaxy at a reasonable distance from this SN, no projected distance could be attributed. The explanation for this is most likely that SN\,2005nc is significantly redshifted because it was accompanied by a gamma-ray burst, GRB\,050525, at a redshift of 0.61. Therefore the host galaxy is most likely too faint to be detected by a two hours exposure of a 2.2\,m telescope.}
\label{SN2005nc}
\end{figure}
Among all of the SNe described in this paper, SN\,2005nc stands out because it is the only SN in the sample that was not selected by a certain distance to its assigned host galaxy but by there simply not being any host galaxy assigned to it in the SAI SN catalog. Since it is instead associated at a short gamma-ray burst, GRB\,050525 with a redshift of $z = 0.61$ \citep{Foley2005}, the task here is to detect the significantly redshifted host galaxy. This does obviously not succeed within the sensitivity limit of our $BV_{\mathrm{Roeser}}$ observations, which in this case is 26.93\,mag\,arcsec$^{-2}$ (see Fig.~\ref{SN2005nc}). Despite SExtractor detecting two point sources close to the cataloged SN position, no galaxy-like source was found during the run configured for detecting extended sources. The final resolution therefore is 0.53$\arcsec$/pixel, which is sufficient for detecting nearby galaxies (as the cut in redshift of z\,$\leq$\,0.05 ensures) but not for galaxies with such a significant redshift where the linear scale rises to 6.75\,kpc/$\arcsec$ (for comparison, at $z = 0.05$ the linear scale is below 1\,kpc/$\arcsec$). Hence, an eventual host galaxy would most likely be a point-like object in our observations for which the detection threshold is $BV_{\mathrm{Roeser}}=24.92$ since these observations were designed to be most sensitive to clearly extended sources in order to detect faint (but nearby) LSB galaxies.

\begin{figure}
\centering
\includegraphics[width=0.5\textwidth]{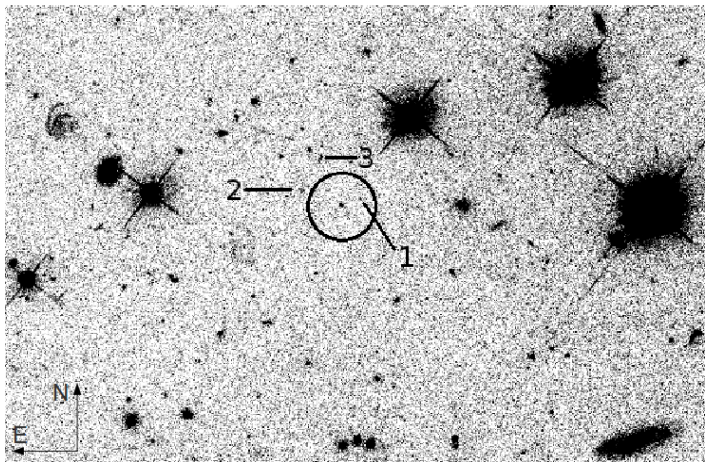}
\caption{R-band ($F625W$) image of the direct vicinity of SN\,2005nc (afterglow of the associated gamma-ray burst GRB\,050525 is located at the center of the circled region which has a radius of 3\arcsec), taken with the Advanced Camera for Surveys onboard the HST on July 30, 66 days after discovery of the GRB. The image's size is $1.0\arcmin\times0.6\arcmin$. Given the point source detection threshold of 27.4\,mag, no host could be identified.}
\label{SN2005nc-ACS}
\end{figure}

Fortunately, SN\,2005nc and the optical afterglow of the associated GRB were observed with the HST's Advanced Camera for Surveys (ACS). Those data have both an excellent spatial resolution (0.05$\arcsec$/pixel) as well as a very high detection limit for point sources of 27.4\,mag. A portion of the corresponding image taken in the $F625W$ filter ($R$-band) and obtained from the {\it Hubble} Legacy Archive \citep[HLA,][]{HLA} is shown in Fig.~\ref{SN2005nc-ACS}. To reach the reported sensitivity limit, we stacked all available $F625W$ images using a sigma clipping algorithm. This caused the sensitivity limit to increase by more than one magnitude (AB) relative to the detection limit given for the individual exposures in the HLA. Comprehensive GRB host galaxy studies carried out with the HST \citep[e.g.][]{Fruchter2006,Fong2010} have shown that the offset between a GRB and its host galaxy can be fairly large, in particular for short-duration GRBs such as 050525. For this class of GRBs, the offset could be as large as 20\,kpc with about 5\% exceeding even this separation \citep{Fong2010}. In the rest frame of GRB\,050525, 20\,kpc is equal to $3\arcsec$, which is the radius of the black circle centered on the optical afterglow and the SN position in Fig.~\ref{SN2005nc-ACS}.
Within this radius, only one point-source could be barely identified by eye (labeled 1 in Fig.~\ref{SN2005nc-ACS}). It is missing in the SExtractor catalog because there are only four connected pixels above the detection threshold of 3$\sigma$, hence we exclude this source from the list of candidate host galaxies and deem it a spurious detection. The nearest galaxy-like object lies 3.5$\arcsec$ to the northeast of the afterglow (labeled 2 in Fig.~\ref{SN2005nc-ACS}), corresponding to 24\,kpc. Although this object is not present in the SExtractor catalog, too, we consider this source to be real after a careful by eye inspection. Nevertheless, because of its large separation from both the SN and the afterglow position, an association with the SN event is unlikely. The nearest SExtractor-identified object is located about 4.5$\arcsec$ to the north of SN\,2005nc (labeled 3 in Fig.~\ref{SN2005nc-ACS}). Despite it also being a very faint object, a slightly extended structure could be found, adding up to a total integrated magnitude of 27.1 as given by SExtractor's AUTOMAG parameter calibrated with a photometric zeropoint derived from the image's header information. At a distance of $z = 0.61$, this converts into a luminosity of $1.6\cdot10^8\,L_{\odot}$, four orders of magnitude less than L$^*$ as derived by \cite{Montero-Dortal2009} for SDSS galaxies. Owing to this extremely low luminosity and the very large separation (more than 30\,kpc) between this object and SN\,2005nc, we also exclude this as the host galaxy. The most reasonable explanation for us is therefore that the faint host galaxy is simply outshone by the slowly dimming supernova itself on that ACS image, which was taken just two months after the discovery of SN\,2005nc. Given the point-source detection threshold of $R$=27.4, we obtain an upper limit of the host galaxy's luminosity of $1.2\cdot10^8\,L_{\odot}$, placing the host of SN\,2005nc at the very end of the SDSS galaxy luminosity distribution by \cite{Montero-Dortal2009}.

\subsection{SN\,2006bx}
\begin{figure}
\centering
\includegraphics[width=0.5\textwidth]{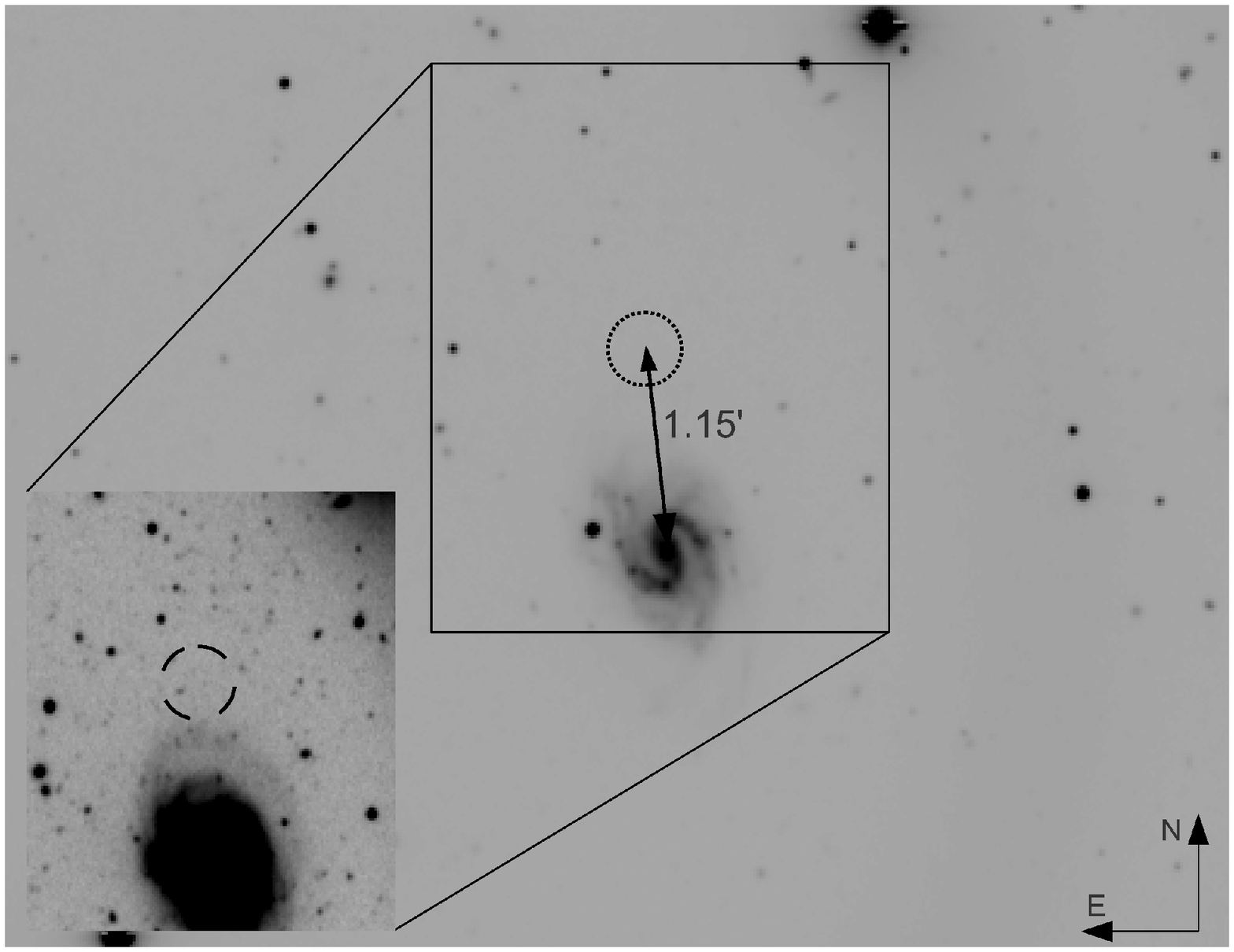}
\label{SN2006bx}
\caption{CAFOS $BV_{\mathrm{Roeser}}$ image of the region around SN\,2006bx, sized $6.7\arcmin\times5.0\arcmin$. See Fig.~\ref{SN1969L} for explanations.}
\end{figure}
SN\,2006bx is a type II supernova associated in the SAI Supernova Catalog with the SAB galaxy UGC\,5434. Its angular separation from the core of UGC\,5434 is 1.20\arcmin, according to a physical distance of $d$ = 26.8\,kpc at the redshift of UGC\,5434 \citep[$z = 0.018613$][]{SDSSDR6}. Since a faint LSB galaxy was not found at the position of SN\,2006bx in our deep $BV_{\mathrm{Roeser}}$ image (whose background is slightly enhanced, a detection threshold of 26.68\,mag\,arcsec$^{-2}$ owing to the reflexion patterns mentioned in Sect.~\ref{obs} was measured) and the GALEX AIS failed to detect any indications of star formation activity at its actual location, we assume that the progenitor of SN\,2006bx is a HVS. This conclusion is also consistent with the literature on UGC\,5434, which does not describe any interactions with other galaxies. Because it is classified as type II supernova with no further characterization, we assume it to be the most common type II explosion, a type IIP. The recent star formation detectable in UGC\,5434 which has a very blue color and significant H\,$\alpha$ emission equivalent to a star formation rate of 0.8$\,M_{\odot}\,\mathrm{yr^{-1}}$ \citep{Kauffmann2003} also makes this explosion type plausible. A type IIP supernova requires a minimum progenitor mass of 8$\,M_{\odot}$ \citep{Smartt2009}, implying a lifetime of shorter than $\tau = 30\,$Myr \citep{Schaller1992}. This gives an estimate for the average velocity of SN\,2006bx's progenitor of
\begin{equation}
v\,\geq\,d/\tau = 848\,\rm{km\,s^{-1}}.
\end{equation}
This falls well within the velocity range applicable to hypervelocity stars, making the progenitor of SN\,2006bx one of the ``fastest'' HVSs known to date. For completeness, we note that UGC\,5434 forms a non-interacting pair with UGC\,05431, an edge-on Scd galaxy located about 6$\arcmin$ to the northwest apart from UGC\,5434 \citep{Nilson1973}. Since there is no evidence that they are interacting in any configuration, in particular because they are separated by more than 23\,Mpc, we disfavor an acceleration scenario for SN\,2006bx's progenitor that involves galaxy -- galaxy interaction. Therefore, the classical acceleration via a close passage of the central SMBH of  SN\,2006bx's progenitor is the most likely explanation of this high velocity.
Assuming a velocity dispersion of 57.9$\pm$5.7\,km\,s$^{-1}$ derived using the SDSS spectrum of UGC\,5434 and the spectral analysis tool of David Schlegel \citep[see also the MPA value added catalog described in][]{Kauffmann2003,Tremonti2004}, one can estimate the central SMBH mass of UGC\,5434. Following the calibration for nearby galaxies of \cite{Gultekin2009}, one comes up with a mass of $\log\left(M_{\rm{SMBH}}/M_{\odot}\right) = 5.84\pm0.75$, nearly one order of magnitude lower than the value for the SMBH at the center of the Milky Way \citep[$\log\left(M_{\rm{SMBH,MW}}/M_{\odot}\right) = 6.42$,][]{Melia2001}. Nevertheless, assuming that the progenitor of SN\,2006bx was a member of a binary system that was accelerated by the collision with the central SMBH of UGC\,5434, the maximum velocity a star could gain through this encounter is about 19,000\,km\,s$^{-1}$ \citep{Tutukov2009}. Hence, the conclusion that SN\,2006bx's progenitor was a HVS is also consistent with current theoretical estimates for the acceleration of HVS.

\section{Summary and conclusions}
We have investigated the phenomenon of host-less supernovae and its origins. A supernova is regarded as host-less if its projected distance $d$ from the nearest possible host galaxy is fairly large, a reasonable cut we established to be $d\,\geq\,10$\,kpc. We searched the Sternberg Astronomical Institute (SAI) supernova catalog, the largest compendium of SN explosions currently available, for host-less core-collapse supernovae defined in this way, finding a small (about 25 SNe) sample for which we have presented deep $BV_{\mathrm{Roeser}}$ observations of a 5 SNe subsample in this paper. Those observations with SExtractor-derived detection thresholds for extended sources of generally 27\,mag\,arcsec$^{-2}$ taken at the 2.2\,m telescope of Calar Alto Observatory were obtained to determine why these SNe were host-less, distinguishing between two explanations already discussed in the literature: (i) The supernova happened in a very faint, potentially low surface brightness (LSB) galaxy that has not been detected within the sensitivity limits of currently available surveys. (ii) The progenitor of the SN has been ejected with an exceptionally high velocity from the galaxy located nearest to it, making it a so-called hypervelocity star (HVS). In addition to these two possibilities, we also verified whether the progenitor had formed {\it in situ} at its actual explosion site using archival GALEX far-UV ($FUV$) data to reveal star formation in the direct vicinity of the supernova. To distinguish between these three explanations, we used our deep $BV_{\mathrm{Roeser}}$ (a very broad filter approximately comprising the Johnson B and V band) observations to search for faint, yet undetected galaxies at the actual SNe sites. If no galaxy could be identified (both by SExtractor and by eye), we discard the option of a faint LSB galaxy being the host of the SNe because the faintest LSB galaxy known to date has a surface brightness of 26.9\,mag\,arcsec$^{-2}$. If there was also no $FUV$ emission detected at the actual SNe sites, we also discarded the possibility of {\it in situ} star formation, leaving a HVS progenitor as the only explanation. Our findings for the individual SNe are listed below:

\begin{enumerate}
\item The progenitor of {\it SN\,1969L}, associated with NGC\,1058 in the SAI SN catalog, is most likely to have formed {\it in situ} because of the very extended $FUV$ disk revealed by GALEX, whose diameter is about three times larger than the optical disk, particularly reaching out to the explosion site of SN\,1969L.
\item {\it SN\,1970L} happened at a projected distance from the center of its associated host galaxy, NGC\,2968, of 14.3\,kpc. This would imply an average projected velocity over the progenitor's lifetime of 1076\,km\,s$^{-1}$. Since no faint LSB galaxy at the actual SN site nor signs of star formation activity could be identified, an acceleration scenario involving the partner of NGC\,2968, NGC\,2970, is possible. Those two galaxies are heavily interacting as is most obvious from the luminous bridge connecting them on which SN\,1970L is also located. Because of the extremely high velocity of more than 1000\,km\,s$^{-1}$ required to reach the SN position, we nevertheless consider a HVS progenitor of SN\,1970L to be unlikely but favor the assumption that it has formed directly within the connecting bridge, despite GALEX being unable to reveal any signs of recent star formation activity. Hence, no final conclusion can be drawn.
\item {\it SN\,1997C} might also be a case in which a tidal interaction between two massive galaxies leads to the acceleration of a star. Since its assigned host galaxy, NGC\,3160, exhibits an S-shaped warp that is commonly thought to originate predominantly from galaxy -- galaxy interactions, this scenario is quite plausible. Owing to its extremely high projected velocity of 3327$\,\rm{km\,s^{-1}}$ that results from following this assumption, emphasize that SN\,1997C was originally misclassified because of the lack of a spectrum obtained around maximum light. Therefore, it might not have been type Ib/c explosion but a Ia with a completely different set of progenitors, in which case our treatment would be inappropriate. Therefore, no definite conclusion can be drawn.
\item {\it SN\,2005nc} is a Ic supernova associated with a gamma-ray burst, GRB\,050525 at $z = 0.61$. It has no host galaxy assigned in the SAI SN catalog, thus our motivation for observing this SN was to detect the highly redshifted host galaxy. This was not possible using our $BV_{\mathrm{Roeser}}$ image, which has a detection limit of 26.96\,mag\,arcsec$^{-2}$ in the direct vicinity of SN\,2005nc. Using archival data obtained with the {\it Hubble} Space Telescope's Advanced Camera for Surveys (ACS) that reaches a point source sensitivity limit of 27.4\,mag, we were similarly unable to detect a host galaxy, most likely because SN\,2005nc was still outshining the host galaxy in these images taken two months after its discovery. Nevertheless, we place upper limits for the host galaxy's luminosity of $1.2\cdot 10^8\,L_{\odot}$, making the host an extremely faint dwarf at the very low-luminosity end of the SDSS galaxy luminosity function.
\item The progenitor of {\it SN\,2006bx} is an excellent example of a HVS that has most likely been accelerated via the classical scenario involving a close passage of the central supermassive black hole  of the parent galaxy (``sling shot manoeuvre''). Because neither signs for either a faint host galaxy or {\it in situ} star formation could be found nor evidence of any interaction with its assigned host galaxy, UGC\,5434, are visible, we favor this classical explanation for the origin of HVS that also applies to most of the HVS in our Galaxy. The projected lifetime-averaged velocity of 848$\,\rm{km\,s^{-1}}$ derived for the progenitor of SN\,2006bx, which lies fairly well within the velocity range of HVS, supports this assumption.
\end{enumerate}

We conclude that the selection criterion for most of our observed sample, that is one based on a projected separation from their assigned host galaxy of more than 10\,kpc, biases the sample toward SNe that have HVS progenitors. This is a fortunate instance because with this one has an excellent instrument to search for HVS in other galaxies than the Milky Way, making the HVS described above the first one observed in a foreign galaxy. Since there are more SNe listed in the SAI SN catalog without any host galaxy assigned to them, we propose to focus on these cases in a second observation run that had already been allocated time. The results will be presented in a subsequent paper. 

\begin{acknowledgements}
We thank our referee, Joel Bregman, and our editor, Ralf Napiwotzki, for their helpful comments on improving the scientific relevance of this paper.
We gratefully thank the Calar Alto staff, in particular Ulli Thiele, for their support in data acquisition and reduction.\\
This research has made use of the NASA/IPAC Extragalactic Database (NED) which is operated by the Jet Propulsion Laboratory, California Institute of Technology, under contract with the National Aeronautics and Space Administration.\\
This research has made use of Aladin \citep{Aladin}.\\
Based on observations made with the NASA/ESA Hubble Space Telescope, and obtained from the Hubble Legacy Archive, which is a collaboration between the Space Telescope Science Institute (STScI/NASA), the Space Telescope European Coordinating Facility (ST-ECF/ESA) and the Canadian Astronomy Data Centre (CADC/NRC/CSA).\\
Funding for the SDSS and SDSS-II has been provided by the Alfred P. Sloan Foundation, the Participating Institutions, the National Science Foundation, the U.S. Department of Energy, the National Aeronautics and Space Administration, the Japanese Monbukagakusho, the Max Planck Society, and the Higher Education Funding Council for England. The SDSS Web Site is http://www.sdss.org/.\\
The SDSS is managed by the Astrophysical Research Consortium for the Participating Institutions. The Participating Institutions are the American Museum of Natural History, Astrophysical Institute Potsdam, University of Basel, University of Cambridge, Case Western Reserve University, University of Chicago, Drexel University, Fermilab, the Institute for Advanced Study, the Japan Participation Group, Johns Hopkins University, the Joint Institute for Nuclear Astrophysics, the Kavli Institute for Particle Astrophysics and Cosmology, the Korean Scientist Group, the Chinese Academy of Sciences (LAMOST), Los Alamos National Laboratory, the Max-Planck-Institute for Astronomy (MPIA), the Max-Planck-Institute for Astrophysics (MPA), New Mexico State University, Ohio State University, University of Pittsburgh, University of Portsmouth, Princeton University, the United States Naval Observatory, and the University of Washington.\\
Some of the data presented in this paper were obtained from the Multimission Archive at the Space Telescope Science Institute (MAST). STScI is operated by the Association of Universities for Research in Astronomy, Inc., under NASA contract NAS5-26555. Support for MAST for non-HST data is provided by the NASA Office of Space Science via grant NNX09AF08G and by other grants and contracts.
\end{acknowledgements}

\bibliographystyle{aa}
\bibliography{HVS_bib}
\end{document}